\documentclass[amsmath,amssymb,amsbsy,twocolumn,superscriptaddress,prb,preprintnumbers,showpacs,footinbib,longbibliography]{revtex4-2}
\usepackage{graphicx,color}
\usepackage{dcolumn}
\usepackage{bm}
\usepackage{braket}
\usepackage{mathtools}
\usepackage{ulem}
\usepackage{mathrsfs}
\usepackage[breaklinks,colorlinks=true,linkcolor=blue,urlcolor=magenta,citecolor=magenta]{hyperref}

\newcommand{\so}[1]{}
\renewcommand{\so}{}
\newcommand{\soo}[1]{}
\newcommand{\sooo}[1]{}

\begin{document}
\title{Orbital-Zeeman cross correlation in $p$- and $d$-wave altermagnets}

\author{Tomonari Mizoguchi}
\affiliation{Department of Physics, University of Tsukuba, 1-1-1 Tennodai, Tsukuba, Ibaraki 305-8571, Japan}
\email{mizoguchi@rhodia.ph.tsukuba.ac.jp}
\author{Soshun Ozaki}
\affiliation{Max Planck Institute for Solid State Research, Heisenbergstrasse 1, D-70569 Stuttgart, Germany}
\date{\today}
\begin{abstract}

Altermagnets are a novel class of magnets
that exhibit a large spin splitting but
the total magnetic moment is vanishing.
This unconventional spin splitting 
gives rise to various characteristic phenomena, such as spin current generation. 
In this paper, 
we study the 
orbital-Zeeman (OZ) cross term in altermagnets. 
Specifically, we consider the Rashba metal and the surface Dirac cones of three-dimensional topological insulators (TIs) in the presence of the altermagnetic order parameters.
For the Rashba metals, 
the $p$-wave order parameter exerts only a limited influence on the OZ term,
whereas the $d$-wave one
causes the sign change of it 
when the order parameter becomes sufficiently large. 
For the TI surface, 
the $p$-wave order parameter
retains the step-function-type dependence of the OZ term as a function of the chemical potential ($\mu$)
associated with the jump at $\mu=0$, observed in the TI surface without magnetism, 
but its magnitude is reduced.
For the $d$-wave case, 
the magnitude of jump at $\mu =0$ is preserved but 
the OZ term decreases as increasing $|\mu|$.
\end{abstract}

\maketitle
\section{Introduction}
Itinerant electron systems with magnetic order exhibit a variety of intriguing physical phenomena. 
Such systems host a delicate interplay between electronic band structures, exchange interactions, and spin-orbit coupling, giving rise to unconventional transport responses and magnetoelectric effects.
Examples include the 
topological and anomalous Hall effects 
in skyrmion systems~\cite{Lee2009,Neubauer2009,Nagaosa2013,Kurumaji2019} and kagome magnets~\cite{Nakatsuji2015,Ikhlas2017,Yang2020}.
Magnetism is also known to 
enrich the topological electronic structures;
antiferromagnetic topological insulators (TIs)~\cite{Mong2010} 
and magnetically doped TIs~\cite{Chang2013,Yoshimi2015} are prominent examples.

In recent years, 
a novel type of magnetic order, dubbed altermagnets~\cite{Naka2019,Smejkal2022,Smejkal2022_2,Hellenes2023,McClarty2024}, 
has attracted considerable attention.
Altermagnets possess properties that are distinct from those of conventional ferromagnets and antiferromagnets.
Specifically, altermagnets exhibit spin splitting even in the absence of spin-orbit coupling, despite having zero net magnetization.
Such a situation typically realizes 
when the opposite-spin sublattices 
are connected by rotation 
but not by translation or inversion. 
Spin splitting in altermagnets has attracted growing interest for its potential in spintronic applications, notably spin current generation~\cite{Naka2019,Naka2021,Smejkal2022,Smejkal2022_2,Naka2025}.%,
Their characteristic 
optical responses~\cite{Ezawa2025_2,Ezawa2025_4}, nonlinear responses~\cite{Fang2024,Ezawa2024_3,Ezawa2025},
the Edelstein effect~\cite{Ezawa2025_3,Chakraborty2025_edel}, 
the peizomagnetic effect~\cite{Naka2025_Piezo,Ogawa2025},
as well as their impacts on other phases of matter
such as 
topological phases~\cite{Ezawa2024_1,Ezawa2024_2} 
and 
superconductivity~\cite{Sumita2023,Chakraborty2024,Sukhachov2024,Mukasa2025,Chakraborty2025,Sumita2025}, have also been actively studied. 
In parallel with theoretical investigations of the characteristic phenomena 
of altermagnets, 
considerable efforts 
have been devoted 
to finding the 
candidate materials 
by \textit{ab initio} calculations~\cite{Noda2016,Ahn2019,Yuan2020,Mazin2021,Ma2021} and 
group-theoretical considerations~\cite{Hayami2019,Hayami2020},
and to experimental detection 
in those materials~\cite{Gonzalez2021,Feng2022,Gonzalez2023,Iguchi2025,Yamada2025}.

In addition to the transport phenomena and optical responses, 
the response to the static magnetic field also reflects 
characteristics of the Bloch electrons.
In fact, the orbital magnetic susceptibility comprises several contributions, including interband processes, a part of which encodes the topology and quantum geometry of the Bloch bands~\cite{Fukuyama1969,Fukuyama1971,Koshino2007,GomezSantos2011,Ogata2015,Gao2015,Piechon2016}.
Additionally, when spin degrees of freedom and orbital motion are coupled,
the magnetic susceptibility acquires additional terms originating from this coupling, which are referred to as the orbital-Zeeman (OZ) cross term~\cite{Murakami2006,Tserkovnyak2015,Suzuura2016,Koshino2016,Nakai2016,Ominato2019,Ozaki2021,Araki2021,Shitade2022,Ozaki2023}.
This cross term also reflects the topological and geometrical aspects of Bloch electrons.
In altermagnets, the magnetic order parameters possess a characteristic momentum dependence, so it is important to clarify how it affects the OZ cross response.
%------------------------------------------------------------------%
\begin{figure}[b]
\begin{center}
\includegraphics[clip,width = \linewidth]{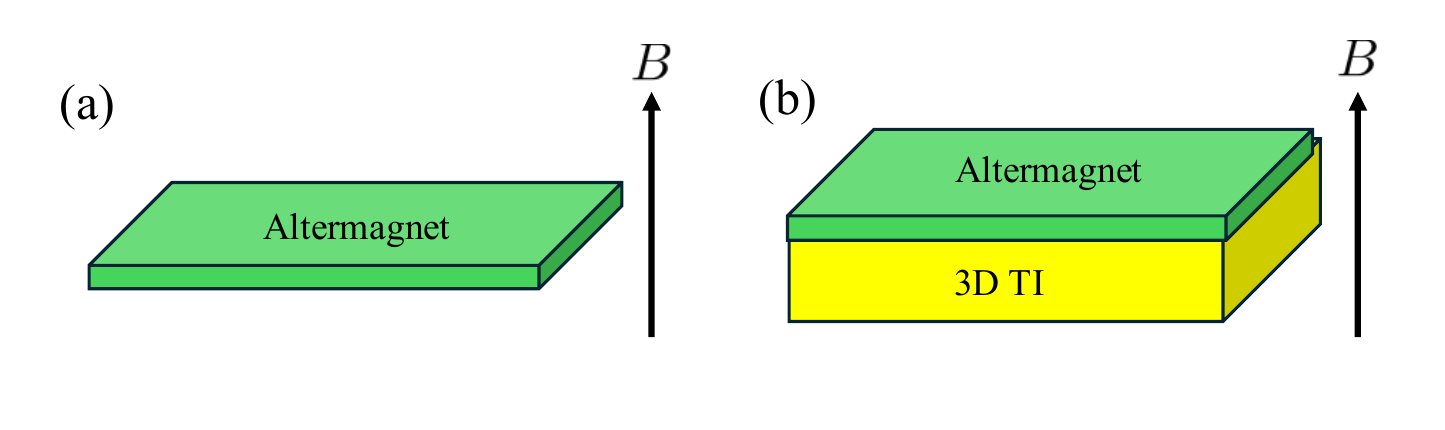}
\vspace{-10pt}
\caption{
Schematic figures of 
(a) the two-dimensional altermagnet
and (b) the altermagnet placed on the three-dimensional TI.
}
\label{fig:setup}
\end{center}
\vspace{-10pt}
\end{figure}
%------------------------------------------------------------------%

In this paper, we study the 
OZ term of the magnetic susceptibility
for the two-dimensional altermagnetic Rashba metal [Fig.~\ref{fig:setup}(a)]
and the surface Dirac cones 
of the three-dimensional TI coupled to the altermagnet [Fig.~\ref{fig:setup}(b)]~\cite{Qin2026}. 
We first point out that 
the altermagnetic order parameter alone cannot induce the OZ term;
rather, the finite non-relativistic effect is necessary.
In this sense, the altermagnetic order plays only a secondary role for the OZ term.
For the Rashba metals, 
we numerically calculate $\chi_{\rm OZ}$
in the low-temperature regime, 
focusing on its chemical potential ($\mu$) dependence.
We find that it can modify the OZ term both qualitatively and quantitatively.
To be specific, 
we find that the $p$-wave order parameter exerts a rather limited influence on the OZ term, whereas the $d$-wave one
causes a sign change of it 
when the order parameter becomes sufficiently large. 
For the TI surface, 
we obtain the analytical expressions 
for $\chi_{\rm OZ}$ at $T=0$,
and find that 
the $p$-wave order parameter
retains the step-function-type dependence of the OZ term as a function of the chemical potential
associated with the jump at $\mu=0$, observed in the TI surface without magnetism, 
but its magnitude is reduced.
For the $d$-wave case, 
the magnitude of jump at $\mu =0$ is preserved but 
the OZ term decreases as increasing $|\mu|$.
In addition, we also remark on the quantization of the jump with respect to the Berry curvature.
%------------------------------------------------------------------%
\begin{figure}[tb]
\begin{center}
\includegraphics[clip,width = \linewidth]{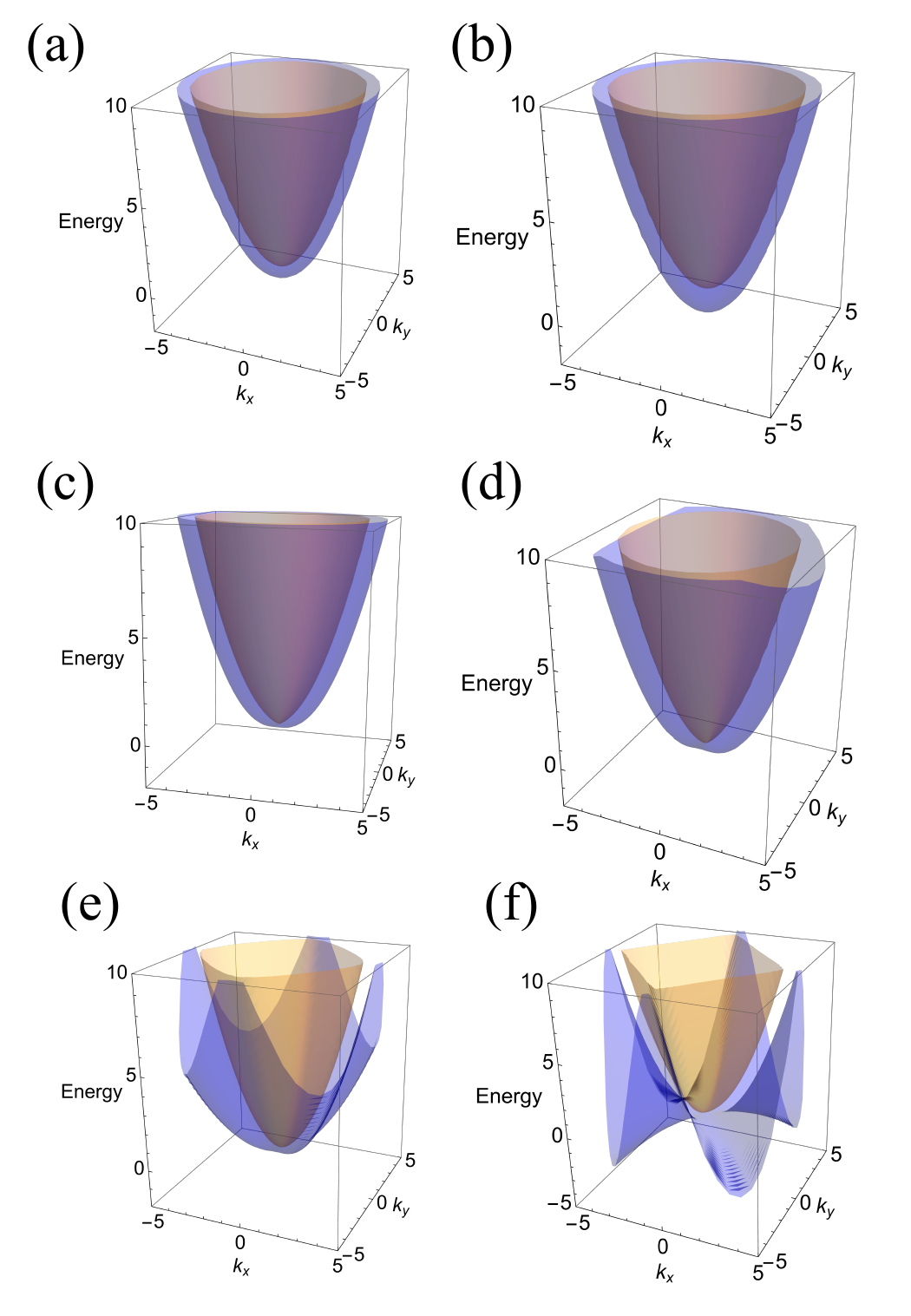}
\vspace{-10pt}
\caption{
Band structures for 
the $s$-wave case with (a) $J = 0.3$ and (b) $J=0.6$,
the $p$-wave case with (c) $J = 0.3$ and (d) $J=0.6$,
and 
the $d$-wave case with (e) $J = 0.3$ and (f) $J=0.6$.
We set $\frac{\hbar^2 }{m} = 1$
and $\lambda = 0.3$.
}
\label{fig:band}
\end{center}
\vspace{-10pt}
\end{figure}
%------------------------------------------------------------------%

The rest of this paper is structured as follows. 
In Sec.~\ref{sec:model}, we introduce the model.
In Sec.~\ref{sec:OZ}, we present the theoretical formulation to calculate $\chi_{\rm OZ}$.
Our main results are presented in Sec.~\ref{sec:results},
where we show the numerical results on the Rashba metal, and the analytic result for the TI surface.
Finally, we summarize this paper in Sec.~\ref{sec:summary}. 

In what follows, 
$\hbar$ represents the reduced Planck constant,
$e$ $(<0)$ represents the electron charge, 
$m$ represents the electron mass,
$\mu_{\rm B}$ represents the Bohr magneton,
and $k_{\rm B}$ stands for the Boltzmann constant. 

\section{Model \label{sec:model}}
We consider the Rashba metal and the surface Dirac cones of TIs in the presence of the altermagnetic order parameters with 
the out-of-plane N\'{e}el vector.
The Hamiltonian in the momentum space reads
\begin{align}
 H_{\bm {k}} = h^0_{\bm{k}} \sigma_0 + \bm{h}_{\bm{k}}\cdot \bm{\sigma},
\end{align}
where $\bm{h}_{\bm{k}}= (h^1_{\bm{k}},h^2_{\bm{k}},h^3_{\bm{k}})$
with $h^1_{\bm{k}} = -\lambda k_y$,
$h^2_{\bm{k}} =  \lambda k_x$,
and $h^3_{\bm{k}} = J X^{\ell}_{\bm{k}}$.
Here, $\sigma_0$ stands for the $2\times 2$ identity matrix and $\bm{\sigma} = (\sigma_x,\sigma_y,\sigma_z)$ are the Pauli matrices that represent the spin degrees of freedom.  
$X_{\bm{k}}^{\ell}$
describes the symmetry of the altermagnet with 
the $\ell$-wave:
\begin{align}
X^\ell_{\bm{k}} = 
\begin{cases}
     1   & \text{ $\ell = s$,}\\
    k_x & \text{$\ell = p$,} \\
    2k_x k_y& \text{$\ell= d$} 
    .
  \end{cases}
\end{align} 
$J$ ($\geq 0$) is the strength of the (alter)magnetic order parameter.
Note that the $s$-wave corresponds to the conventional ferromagnet,
which we consider for the purpose of comparison with the altermagnet.
As for $h^0_{\bm{k}}$, when considering the Rashba metal, we set
$h^0_{\bm{k}} = \frac{\hbar^2 k^2}{2m}$ ($k = \sqrt{k_x^2  + k_y^2}$).
In that case, $\lambda$ stands for 
the coupling constant of the Rashba spin-orbit coupling. 
For the TI surface, we set 
$h^0_{\bm{k}} =0$, 
and in that case 
$\lambda$ corresponds to 
$\lambda = \hbar v_{\rm F}$
where $v_{\rm F}$ is the Fermi velocity of the Dirac cone. 

In what follows, the band structure plays an important role in discussing $\chi_{\mathrm{OZ}}$ of Rashba metals.
In Fig.~\ref{fig:band}, we plot 
the band structures for 
the $s$-, $p$-, and $d$-wave cases.
We see two key features. 
First, the upper and the lower bands are gapped for the $s$-wave case, whereas the gap closes at $\bm{k} = \bm{0}$ for $p$- and $d$-wave cases,
which is attributed to the nodal structure of the order parameters. 
Second, only in the $d$-wave case, the lower band has no lower bound when $J$ is sufficiently large.
Specifically, we find that a band structure unbounded from below is realized when $J > \frac{\hbar^2}{2m}$.

\section{Orbital-Zeeman cross correlation \label{sec:OZ}}
We calculate 
the magnetic susceptibility 
for the cross correlation between the orbital and Zeeman effects, 
$\chi_{\rm OZ}$, 
whose microscopic derivation has been given in Refs.~\cite{Suzuura2016,Nakai2016,Ozaki2021,Ozaki2023}.
We consider the case where
the magnetic field is in the $z$-direction, which is perpendicular to the two-dimensional plane and is parallel to the N\'{e}el vector.
In this case, $\chi_{\rm OZ}$ is given as
\begin{align}
\chi_{\rm OZ} 
= \frac{i|e|\mu_{\rm B}}{\hbar}
\frac{k_{\rm B} T}{V}
\sum_{n, \bm{k}} 
\mathrm{Tr}
\left[ 
\sigma^3 \mathcal{G} \gamma_x 
\mathcal{G} \gamma_y
\mathcal{G}
- (x \leftrightarrow y)
\right], \label{eq:chiOZ}
\end{align}
where 
\begin{align}
\mathcal{G} 
=&  (i\omega_n +\mu -H_{\bm{k}})^{-1}  
= \frac{1}{(i\omega_n - \xi_{\bm{k},+})(i\omega_n- \xi_{\bm{k},-})}\notag \\
\times&\begin{pmatrix}
i\omega_n +\mu-h^0_{\bm{k}} + h^3_{\bm{k}} & h^1_{\bm{k}} - i h^2_{\bm{k}} \\
h^1_{\bm{k}} + i h^2_{\bm{k}} &
i\omega_n +\mu-h^0_{\bm{k}} - h^3_{\bm{k}} \\
\end{pmatrix} \label{eq:GF}
\end{align}
is the Matsubara Green's function with $\omega_n = (2n+1)\pi k_{\rm B}T$,
$\gamma_\nu = \partial_{k_\nu} H_{\bm{k}}$
is the velocity operator in $\nu$-direction multiplied by $\hbar$, and
$V$ is the area of the two dimensional system.
In Eq.~(\ref{eq:GF}), we have used $\xi_{\bm{k},\pm}
= h^0_{\bm{k} } \pm h_{\bm{k}} 
-\mu$,
with $h_{\bm{k}}:=|\bm{h}_{\bm{k}}|$.
For later use, we further define $\Omega_{\bm{k}}(i\omega_n) :=i\omega_n +\mu-h^0_{\bm{k}}
= i\omega_n -\xi^0_{\bm{k}}$,
with $\xi^0_{\bm{k}}:= h^0_{\bm{k}}-\mu$. 

Taking the trace in Eq.~(\ref{eq:chiOZ}), 
we have
\begin{align}
&\mathrm{Tr}\left[
\sigma^3 \mathcal{G} \gamma_x 
\mathcal{G} \gamma_y
\mathcal{G}
- (x \leftrightarrow y)
\right] \notag \\
= & \frac{4i \lambda^2 }{(i\omega_n - \xi_{\bm{k},+})^3(i\omega_n- \xi_{\bm{k},-})^3}
\sum_{j=0}^3 P^{\ell}_{\bm{k},j}
\Omega^j, \label{eq:matsu_sum}
\end{align}
where $P^{\ell}_{\bm{k},j}$'s are given as
\begin{subequations}
\begin{align}
    P^{\ell}_{\bm{k},0}
    = Y \cdot -\frac{\hbar^2 k^2}{m}h^2_{\bm{k}},
\end{align}
\begin{align}
    P^{\ell}_{\bm{k},1}
    = -h^2_{\bm{k}}
\end{align}
\begin{align}
    P^{\ell}_{\bm{k},2}
    = Y \cdot \hbar^2 \frac{k^2}{m}
\end{align}
and
\begin{align}
    P^{\ell}_{\bm{k},3}
    = 1,
\end{align}
\end{subequations}
where $Y = 1 (0)$ for the Rashba metal (the TI surface).
Note that $P^\ell_{\bm{k},2}$ and 
$P^\ell_{\bm{k},3}$ do not depend on $\ell$.
In Eq.~(\ref{eq:matsu_sum}),
we see 
$\chi_{\rm OZ} = 0$
if $\lambda = 0$,
which means that the altermagnet alone cannot induce the OZ cross response.
In fact, the spin is the good quantum number for $\lambda = 0$,
meaning that $\mathcal{G}$ and $\gamma_{j}$
are the diagonal matrices.
Therefore, we have
$ \sigma^3\mathcal{G}\gamma_x\mathcal{G}\gamma_y \mathcal{G}  = \sigma^3\mathcal{G}\gamma_y\mathcal{G}\gamma_x \mathcal{G}$ as all the matrices commute with each other,
resulting in the vanishing of $\chi_{\rm OZ}$. 
%------------------------------------------------------------------%
\begin{figure*}[tb]
\begin{center}
\includegraphics[clip,width = \linewidth]{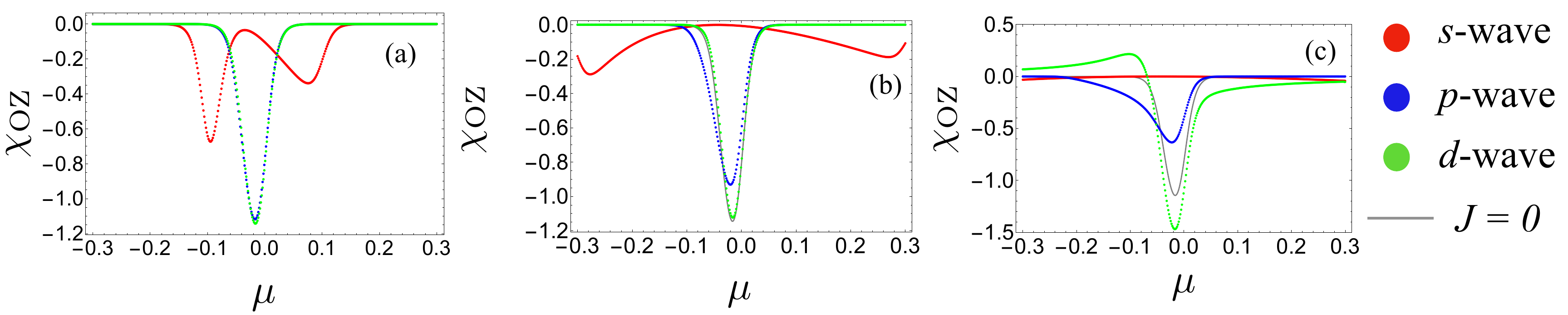}
\vspace{-10pt}
\caption{
$\mu$ dependence of $\chi_{\rm OZ}$
for the Rashba metal
for $\lambda = 0.3$
and 
(a) $J= 0.1$, 
(b) $J=0.3$,
and (c) $J=0.6$.
The unit of the vertical axis is $\frac{|e|\mu_{\rm B}}{2\pi\hbar}$.
}
\label{fig:chioz_r}
\end{center}
\vspace{-10pt}
\end{figure*}
%------------------------------------------------------------------%

To proceed further, we need to 
perform the summation over the Matsubara frequency.
After some algebras 
(see Appendix~\ref{app:matsubara}),
we have the following expression:
\begin{align}
& \chi_{\rm OZ} \notag \\ 
=&\frac{-4\lambda^2|e|\mu_{\rm B}}{\hbar}
\int \frac{d^2k}{(2\pi)^2} 
\left[\sum_{\alpha = \pm}
G^{\ell,(0)}_{\bm{k},\alpha} f(\xi_{\bm{k},\alpha})
+ G^{\ell,(1)}_{\bm{k},\alpha} f^\prime (\xi_{\bm{k},\alpha})
\right],\label{eq:chiOZ_final}
\end{align}
where
\begin{subequations}
\begin{align}
G^{\ell,(0)}_{\bm{k},\alpha}
= \alpha\left[\frac{3}{16h_{\bm{k}}^5} P^{\ell}_{\bm{k},0} - \frac{1}{16h_{\bm{k}}^3} P^{\ell}_{\bm{k},2}\right]
= -\frac{Y \alpha}{4 h_{\bm{k}}^3} \frac{\hbar^2 k^2}{m},
\end{align}
and
\begin{align}
& G^{\ell,(1)}_{\bm{k},\alpha}\notag \\
=& -\frac{3}{16h_{\bm{k}}^4} P^{\ell}_{\bm{k},0} 
-\alpha \frac{1}{16h_{\bm{k}}^3} P^{\ell}_{\bm{k},1}
+ \frac{1}{16h_{\bm{k}}^2} P^{\ell}_{\bm{k},2}
+ \alpha \frac{3}{16h_{\bm{k}}} P^{\ell}_{\bm{k},3} \notag \\
=& 
\frac{\alpha}{4h_{\bm{k}}} + 
\frac{Y}{4h_{\bm{k}}^2}
\frac{\hbar^2 k^2}{m},
\end{align}
\end{subequations}
where
$f(x) = \frac{1}{e^{x/k_{\rm B}T} + 1}$
is the Fermi-Dirac distribution function
and $f^\prime(x)$ is its derivative.
It is to be noted that the term proportional to $f$ vanishes for the TI surface.
Without the altermagnet,
(i.e., $J=0$), 
analytic results are obtained for several cases,
which correspond to the results of the previous works~\cite{Tserkovnyak2015,Suzuura2016}.
We summarize those results in Appendices~\ref{app:rashba} and \ref{app:TI}. 
Note also that, for the TI surface,
we derive the explicit forms 
of $\chi_{\rm OZ}$ 
at any temperature 
for the $s$- and $p$-wave cases,
and at $T=0$ for the $d$-wave
case.
These results are also presented in Appendix~\ref{app:TI}.

\section{Results \label{sec:results}}
%------------------------------------------------------------------%
\begin{figure}[b]
\begin{center}
\includegraphics[clip,width = 0.95\linewidth]{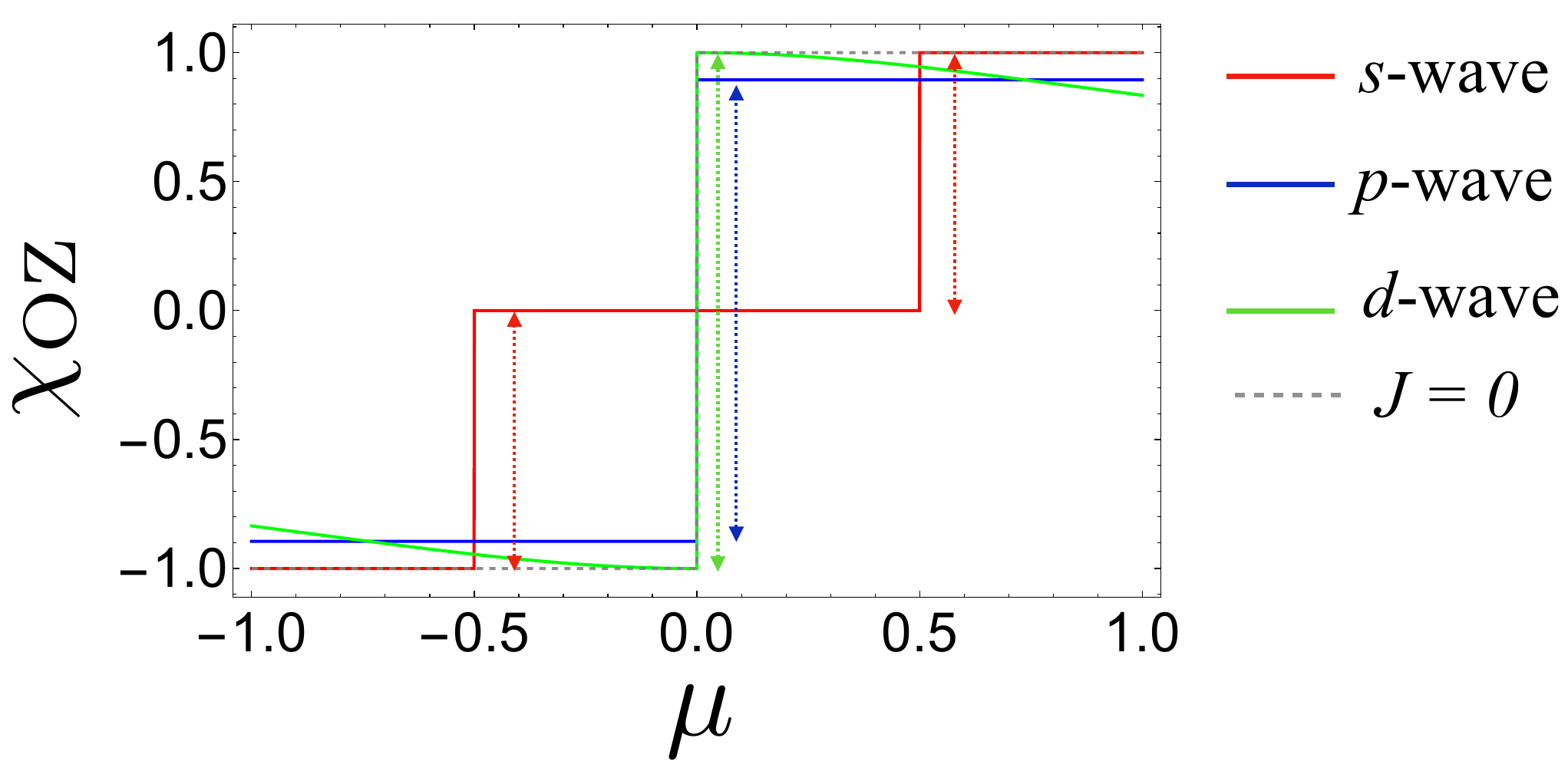}
\vspace{-10pt}
\caption{$\mu$ dependence of $\chi_{\rm OZ}$ 
for the TI surface
with $\lambda = 1$ and $T=0$.
The unit of the vertical axis is $\frac{|e|\mu_{\rm B}}{2\pi\hbar}$.}
\label{fig:chioz_ti}
\end{center}
\vspace{-10pt}
\end{figure}
%------------------------------------------------------------------%
\subsection{Rashba metal}
For the Rashba metal, we perform the integration over $\bm{k}$ numerically. 
We set $k_{\rm B} T= 0.01$ and $\frac{\hbar^2}{m} = 1$. 
For the integration over $\bm{k}$,
we employ the radial coordinate
$\int\frac{d^2k}{(2\pi)^2} \rightarrow 
\frac{1}{{(2\pi)^2}}
\int_0^{k_c} k dk \int_{0}^{2\pi} d\theta$,
where $(k_x,k_y) = (k\cos \theta, k\sin \theta)$ and
$k_c$ is the momentum cut-off.
Here we set $k_c = 6\pi$,
 which is far away from the Fermi surface except for the case of the $d$-wave with large $J$.

Figure~\ref{fig:chioz_r} shows the $\mu$ dependence of $\chi_{\rm OZ}$
for the Rashba metal with $\lambda = 0.3$.
For comparison, we also plot the case of $J=0$, which exhibits 
the diamagnetic susceptibility 
around $\mu =0$.
For the $s$-wave, 
which opens the band gap at $\bm{k}=0$,
$\chi_{\rm OZ}$ is suppressed when $\mu$ lies in that gap 
[i.e., the case where only one Fermi surface remains
as shown in Figs.~\ref{fig:band}(a) and \ref{fig:band}(b)],
and it has minima at the edge of that gap. 
In the $p$-wave case, 
the diamagnetic susceptibility is monotonically suppressed as $J$ increases,
which indicates that the $p$-wave order parameter
affects the OZ term only in a quantitative manner.
In contrast to the $p$-wave,
we see a prominent change for $d$-wave. 
Namely, for sufficiently large $J$ [i.e., $J=0.6$ in Fig.~\ref{fig:chioz_r}(c)],
we see that $\chi_{\rm OZ}$ becomes positive 
for $\mu \lesssim -0.069$.
This paramagnetic  $\chi_{\mathrm{OZ}}$ may be attributed to the unbounded lower band [Fig.~\ref{fig:band}(f)].
It then turns negative,
and the diamagnetic susceptibility at the minimum of $\chi_{\rm OZ}$ is enhanced compared with the $J=0$ case.

\subsection{Topological insulator surface}
As we have mentioned in the previous section, 
for the TI surface,
we can perform the integration over $\bm{k}$ analytically. 
The detailed derivation is given in Appendix~\ref{app:TI}; here we summarize the results for $T=0$.
For the $s$-wave magnet, 
we have 
\begin{align}
\chi_{\rm OZ}
= \frac{|e|\mu_{\rm B}}{2\pi \hbar}[\Theta(\mu-J)+\Theta(J+\mu)- 1],
\label{eq:chioz_s_t0}
\end{align}
for the $p$-wave magnet, 
we have 
\begin{align}
\chi_{\rm OZ}
= \mathrm{sgn}(\mu)\frac{|e|\mu_{\rm B}}{2\pi \hbar}\frac{\lambda}{\sqrt{\lambda^2 + J^2}},
\end{align}
and for the $d$-wave magnet, 
we have 
\begin{align}
\chi_{\rm OZ}
=\mathrm{sgn}(\mu) \frac{|e|\mu_{\rm B}}{\hbar}
\frac{K(w)}{\pi^2\sqrt{1+w}},
\end{align} 
where $\Theta (x)$ represents the step function,
$w = \frac{4J^2 \mu^2}{\lambda^4}$, and $K(x)$ is the complete elliptic integral of the first kind.
Note that for $J=0$,
i.e., the TI surface without magnetism, we have
\begin{align}
\chi_{\rm OZ}
= \mathrm{sgn}(\mu) 
\frac{|e|\mu_{\rm B}}{2\pi \hbar}.\label{eq:chioz_j0_t0}
\end{align}
Note also that the results for the $s$-wave magnet as well as $J=0$ as a limit of that were obtained by Ref.~\onlinecite{Tserkovnyak2015}.

In Fig.~\ref{fig:chioz_ti},
we plot the $\mu$ dependence of
$\chi_{\rm OZ}$ at $T=0$ with $\lambda = 1$
and $J=0.5$;
for comparison, we also plot the case of $J=0$.
We see a jump of $\chi_{\rm OZ}$ at $\mu = 0$ for $J=0$,
the $p$-wave, and 
the $d$-wave cases,
and $\mu = \pm J$ for 
the $s$-wave case. 
As for the magnitude of the jump, we see that 
$\Delta \chi_{\rm OZ} 
:= \frac{|e|\mu_{\rm B}}{\pi \hbar}$
for $J=0$ and the $d$-wave case,
which is the universal value,
whereas
$\Delta \chi_{\rm OZ} 
:= \frac{\lambda}{\sqrt{\lambda^2+J^2}}\frac{|e|\mu_{\rm B}}{\pi \hbar}$ for the $p$-wave,
which is parameter dependent.
For the $s$-wave, we see two jumps each of which has half of the magnitude of the $J=0$ value.

In addition to the jump, we also find that, except for the $d$-wave case, 
$\chi_{\rm OZ}$ exhibits the step-function $\mu$ dependence,
namley, $\chi_{\rm OZ} \propto \mathrm{sgn}(\mu)$ for $J=0$ and the 
$p$-wave, and $\chi_{\rm OZ} \propto [\Theta(\mu-J)+\Theta(J+\mu)- 1]$ for the $s$-wave.
For the $d$-wave, on the other hand, 
$|\chi_{\rm OZ}|$ decreases as  increasing $|\mu|$.

\so{Finally, we address the origin of the quantized jump width in $\chi_{\rm OZ}$ observed for the $J=0$ and $s$-wave order parameters, both of which have particle-hole symmetry. Half of the contribution is interpreted as the orbital magnetization induced by the Zeeman field \cite{Murakami2006, Nakai2016, Ozaki2021,Ozaki2023}:
\soo{
\begin{equation}
    \frac{1}{2}\chi_{\rm OZ}=\frac{\partial M_z(B)}{\partial B} \biggr|_{B=0},
\end{equation}
where $M_z(B)$ is the orbital magnetization in the $z$ direction, and $B$ is the magnetic field coupled to spin via the Zeeman term.
}
For our two-band systems with particle-hole symmetry, the orbital magnetization at zero temperature is expressed as \cite{Xiao2005,Thonhauser2005,Xiao2007,Souza2008,Ogata2017}
\begin{equation}
    M_z(B)=\frac{|e|\mu}{\hbar} \sum_{ {\bm k},\pm} \Omega^\pm_{\bm z}({\bm k})\Theta(\mu-E^\pm_{\bm k}(B)),
    \label{eq:Mz}
\end{equation}
where $E^\pm_{\bm k}(B)$ and $\Omega^\pm_z(\bm k)$ represent the energy dispersion of the conduction/valence band and its Berry curvature, respectively.

\soo{This calculation reproduces the results for $J=0$ and $s$-wave cases at $T=0$. The details of the calculation are presented in Appendix~\ref{app:TI_interp}.}
These expressions indicate that the Berry-curvature effect leads to the quantization of the jump, which is robust against spin-orbit coupling. We expect that the same mechanism is responsible for the $d$-wave order parameter case, which also possesses particle-hole symmetry.
\soo{The breaking of the quantization in the $p$-wave case is attributed to the absence of the 
particle-hole symmetry, which does not allow for the expression in Eq.~\eqref{eq:Mz}.}
} 

\section{Summary \label{sec:summary}}
We have studied the orbital-Zeeman cross correlation 
of the altermagnets.
We have considered the two cases, namely,
the altermagnet in the presence of the Rashba spin-orbit coupling 
and the altermagnet coupled to 
the surface Dirac cone of the three-dimensional TI. 
Although the effect of the altermagnetic order parameter is limited to be secondary,
we find several characteristic features.
For the Rashba metal, 
we find that the $p$-wave order parameter causes only quantitative changes in $\chi_{\rm OZ}$,
whereas the $d$-wave can cause 
the qualitative change, 
such as the sign change,
if the order parameter is large enough. 
For the TI surface, 
we obtain the analytical expressions 
for $\chi_{\rm OZ}$.
To be precise, 
we obtain the full chemical potential and temperature dependence for the $p$-wave case,
whereas for 
the $d$-wave case the analytic form is limited to the zero temperature.
Focusing on the chemical potential dependence 
at the zero temperature, we find that 
the $p$-wave order parameter
retains the step-function-type dependence of the OZ term as a function of the chemical potential
associated with the jump at $\mu=0$, observed in the TI surface without magnetism, 
but its magnitude becomes parameter dependent rather than the universal value.
For the $d$-wave case, 
the jump at $\mu =0$ is the universal value as is the case with $J=0$, 
but the OZ term decreases 
as increasing $|\mu|$, which is dictated by the complete elliptic integral of the first
kind.
We have also elucidated  that the quantization of the jump for the $s$-wave case originates from the Berry curvature effect.

\acknowledgements
We thank Masao Ogata,
Nobuyuki Okuma, and Ikuma Tateishi 
for fruitful discussions.
This work is supported by 
JSPS KAKENHI, Grant 
No.~JP23K03243 (TM).

\appendix
\section{Derivation of Eq.~(\ref{eq:chiOZ_final}) \label{app:matsubara}}
Here we describe the details
of the summation over Matsubara frequency in Eq.~(\ref{eq:matsu_sum}).
As is the standard method 
using the residue integral, we have
\begin{align}
&k_{\rm B} T \sum_n
\frac{\Omega^j}{(i\omega_n - \xi_{\bm{k},+})^3(i\omega_n- \xi_{\bm{k},-})^3} \notag \\
=& \frac{1}{2}   
   [Q^j(\xi_+; \xi_-) + Q^j(\xi_-; \xi_+)],
\end{align}
where
\begin{align}
Q^j(x;y) = \frac{d^2}{dx^2} \left[\frac{f(x) (\Omega_{\bm{k}} (x))^j}{(x-y)^3} \right].
\end{align}
The explicit forms of $Q^j(x;y)$ are 
\begin{subequations}
\begin{align}
Q^0(x;y) = 
\frac{f^{\prime \prime} (x)}{(x-y)^3}-\frac{6 f^\prime(x)}{(x-y)^4}+\frac{12 f(x)}{(x-y)^5},
\end{align}
\begin{align}
Q^1(x;y) =& 
\frac{(x-\xi^0)}{(x-y)^3}f^{\prime \prime} (x)
+ \left[
\frac{2}{(x-y)^3}
-\frac{6(x-\xi^0)}{(x-y)^4}
\right]
f^{\prime} (x) \notag \\
+& \left[
\frac{12(x-\xi^0)}{(x-y)^5}
-\frac{6}{(x-y)^4}
\right]
f(x),
\end{align}
\begin{align}
Q^2(x;y) =& 
\frac{(x-\xi^0)^2}{(x-y)^3}
f^{\prime \prime}(x)
+ \left[
\frac{4(x-\xi^0)}{(x-y)^3}
-\frac{6(x-\xi^0)^2}{(x-y)^4}
\right]
f^{\prime} (x)\notag \\
+& \left[
\frac{12(x-\xi^0)^2}{(x-y)^5}
-\frac{12(x-\xi^0)}{(x-y)^4}
+\frac{2}{(x-y)^3}
\right]
f(x),
\end{align}
and 
\begin{align}
Q^3(x;y) =&
\frac{(x-\xi^0)^3}{(x-y)^3}
f^{\prime \prime}(x)
+ \left[
\frac{6(x-\xi^0)^2}{(x-y)^3}
-\frac{6(x-\xi^0)^3}{(x-y)^4}
\right]
f^{\prime} (x)\notag \\
+& \left[
\frac{12(x-\xi^0)^3}{(x-y)^5}
-\frac{18(x-\xi^0)^2}{(x-y)^4} 
+\frac{6(x-\xi^0)}{(x-y)^3}
\right]
f(x).
\end{align}
\end{subequations}
Using these, we have
\begin{subequations}
\begin{align}
&k_{\rm B} T \sum_n
\frac{1}{(i\omega_n - \xi_{\bm{k},+})^3(i\omega_n- \xi_{\bm{k},-})^3} \notag \\
=& \frac{3[f(\xi_+)-f(\xi_-)]}{16 h^5}
- \frac{3[f^\prime(\xi_+)+f^\prime(\xi_-)]}{16 h^4}
+ \frac{f^{\prime \prime}(\xi_+)- f^{\prime \prime} (\xi_-)}{16 h^3},
\end{align}
\begin{align}
&k_{\rm B} T \sum_n
\frac{\Omega}{(i\omega_n - \xi_{\bm{k},+})^3(i\omega_n- \xi_{\bm{k},-})^3} \notag \\
&= 
- \frac{f^\prime(\xi_+)-f^\prime(\xi_-)}{16 h^3}
+ \frac{f^{\prime \prime}(\xi_+)+ f^{\prime \prime} (\xi_-)}{16 h^2},
\end{align}
\begin{align}
&k_{\rm B} T \sum_n
\frac{\Omega^2}{(i\omega_n - \xi_{\bm{k},+})^3(i\omega_n- \xi_{\bm{k},-})^3} \notag \\
=& - \frac{[f(\xi_+)-f(\xi_-)]}{16 h^3}
+ \frac{f^\prime(\xi_+)+f^\prime(\xi_-)}{16 h^2}
+ \frac{f^{\prime \prime}(\xi_+)- f^{\prime \prime} (\xi_-)}{16 h},
\end{align}
and
\begin{align}
&k_{\rm B} T \sum_n
\frac{\Omega^3}{(i\omega_n - \xi_{\bm{k},+})^3(i\omega_n- \xi_{\bm{k},-})^3} \notag \\
=&\frac{3[f^\prime(\xi_+)-f^\prime(\xi_-)]}{16 h}
+ \frac{f^{\prime \prime}(\xi_+)+  f^{\prime \prime} (\xi_-)}{16}.
\end{align}
\end{subequations}
Here, we use the shorthand $h$ to denote $h_{\bm{k}}$.
Substituting these into
Eq.~(\ref{eq:matsu_sum}), we obtain 
the form of Eq.~(\ref{eq:chiOZ_final}).
Notice that the term proportional to 
$f^{\prime \prime}$ vanishes for both the Rashba metal and the TI surface.

\section{Analytic results for Rashba metal \label{app:rashba}}
%------------------------------------------------------------------%
\begin{figure}[t]
\begin{center}
\includegraphics[clip,width = 0.95\linewidth]{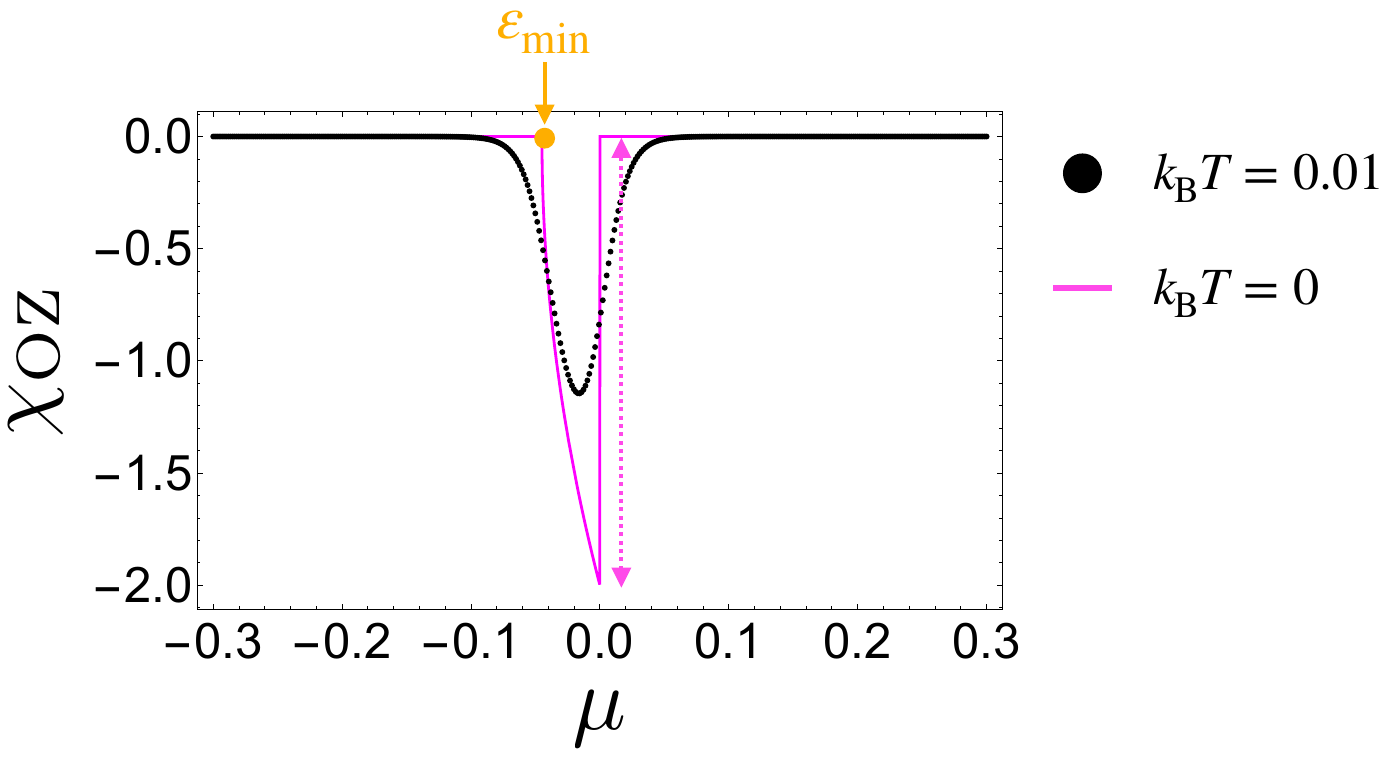}
\vspace{-10pt}
\caption{$\mu$ 
dependence of 
$\chi_{\rm OZ}$ 
for the Rashba metal 
with $\lambda = 0.3$.
The black dots represent the numerical result with $k_{\rm B} T = 0.01$
and the magenta line represent 
the result for 
$k_{\rm B} T = 0$ [Eq.~(\ref{eq:chioz_rashba})].
The unit of the vertical axis is $\frac{|e|\mu_{\rm B}}{2\pi\hbar}$.}
\label{fig:chioz_rashba_exact}
\end{center}
\vspace{-10pt}
\end{figure}
%------------------------------------------------------------------%
In this appendix, we show the analytic result on the Rashba metal without 
the altermagnetic order parameter,
which was obtained by Ref.~\onlinecite{Suzuura2016}.
In this case,
we have
\begin{align}
G^{(0)}_{\alpha} = -\frac{\alpha \hbar^2}{4m \lambda^3 k},
\end{align}
and
\begin{align}
G^{(1)}_{\alpha} = 
\frac{\hbar^2}{4m \lambda^2}
+
\frac{\alpha}{4 \lambda k}.
\end{align}
Thus, we have
\begin{widetext}
\begin{align}
\chi_{\rm OZ}
=& -\frac{\lambda^2|e| \mu_{\rm B}}{2\pi \hbar}\sum_{\alpha=\pm} 
\int_0^{\infty}dk
\left[
-\frac{\alpha \hbar^2}{m\lambda^3}f(\xi_\alpha)
+ \left( 
\frac{\alpha}{\lambda} + \frac{\hbar^2 k}{m\lambda^2} \right)f^\prime(\xi_{\alpha})
\right] 
= -\frac{\lambda^2|e| \mu_{\rm B}}{2\pi \hbar}\sum_{\alpha=\pm} 
\int_0^{\infty}dk
\left[
-\frac{\alpha \hbar^2}{m\lambda^3}f(\xi_\alpha)
+ \frac{1}{\lambda^2} \frac{d f(\xi_{\alpha})}{dk}
\right].
\end{align}
\end{widetext}
Note that we have used
\begin{align}
\frac{d\xi_\alpha}{dk}
= \frac{\hbar^2 k}{m} +\alpha \lambda
= \lambda^2 \left(
\frac{\alpha}{\lambda} + \frac{\hbar^2 k}{m\lambda^2}
\right).
\end{align}

To proceed further, we consider the case of $T=0$.
We first consider the case of $\mu > 0$.
In this case, we have
\begin{align}
\int_0^\infty dk f(\xi_{\alpha})
= k_{\rm F, \alpha},
\end{align}
where $k_{\rm F, \alpha}$ satisfies
$\xi_{k_{\rm{F},\alpha},\alpha} = 0$.
To be specific, 
they are given as
\begin{align}
k_{\rm{F},\alpha} = \frac{m}{\hbar^2}\left(-\alpha \lambda + \sqrt{\lambda^2 + 2 \mu \frac{\hbar^2}{m}}\right).
\end{align}
We also have 
\begin{align}
\int_0^\infty dk  \frac{d f(\xi_{\alpha})}{d k}
= [\Theta(-\xi_{k,\alpha})]_0^{\infty}
= -1.
\end{align}
Substituting these, 
we have 
\begin{align}
\chi_{\rm OZ}(T=0, \mu > 0)
=& -\frac{\lambda^2|e| \mu_{\rm B}}{2\pi \hbar}
\left[\frac{\hbar^2}{m\lambda^3}
(k_{\rm{F},-}-k_{\rm{F},+})
- \frac{2}{\lambda^2}
\right] \notag \\
=& 0.
\end{align}
Note that we have used 
$k_{\rm{F},-} - k_{\rm{F},+} = \frac{2 m \lambda}{\hbar^2}$.

For $\mu < 0$, only the band $\alpha = -$ gives the finite contribution.
In this case, we have 
\begin{align}
\int_0^\infty dk f(\xi_{-})
= \left(k^{(1)}_{\rm F, -} - k^{(2)}_{\rm F, -}\right)\Theta(\mu-\varepsilon_{\rm min}),
\end{align}
where $\varepsilon_{\rm min}= -\frac{\lambda^2 m}{2\hbar^2}$ is the bottom of the lower band, and
\begin{align}
k^{(1)}_{\rm F, -} =& \frac{m}{\hbar^2} \left(\lambda +\sqrt{\lambda^2 + 2 \mu \frac{\hbar^2}{m}} \right), \notag \\
k^{(2)}_{\rm F, -} =& \frac{m}{\hbar^2} \left(\lambda - \sqrt{\lambda^2 + 2 \mu \frac{\hbar^2}{m}} \right).
\end{align}
We also have
\begin{align}
\int_0^\infty dk  \frac{d f(\xi_{-})}{d k}
= [\Theta(-\xi_{k,-})]_0^{\infty}
= 0.
\end{align}
Then, we have
\begin{align}
\chi_{\rm OZ}(T=0, \mu < 0)
= -\Theta(\mu-\varepsilon_{\rm min})\frac{|e| \mu_{\rm B}}{2\pi \hbar}
\left[
 \frac{2\sqrt{\lambda^2 + 2 \mu \frac{\hbar^2}{m}}}{\lambda}
\right].
\end{align}
Combining these, we have
\begin{align}
\chi_{\rm OZ}(T=0)
= -\frac{|e| \mu_{\rm B}}{2\pi \hbar}
\left[
 \frac{2\sqrt{\lambda^2 + 2 \mu \frac{\hbar^2}{m}}}{\lambda}
\right]
\Theta(-\mu)\Theta(\mu-\varepsilon_{\rm min}). \label{eq:chioz_rashba}
\end{align}
Note that $\chi_{\rm OZ}$
exhibits a jump at 
$\mu = 0$, where $\chi_{\rm OZ}$ is minimized,
of magnitude 
$\frac{|e|\mu_{\rm B}}{\pi \hbar}$, which is 
independent 
of the parameters.

In Fig.~\ref{fig:chioz_rashba_exact},
we plot $\chi_{\rm OZ}$ of Eq.~(\ref{eq:chioz_rashba}) for $\lambda = 0.3$,
together with the numerical result for $k_{\rm B} T = 0.01$.
Comparing these two cases,
we see that the minimum of $\chi_{\rm OZ}$ is shifted to the finite $\mu$ $(<0)$ 
and the value of the minimum is largely reduced
for $k_{\rm B} T = 0.01$.

\section{Analytic results for topological insulator surface \label{app:TI}}
In this appendix, we show the derivations of Eqs.~(\ref{eq:chioz_s_t0})-(\ref{eq:chioz_j0_t0}), that are, 
the analytic results on the TI surface.
Note again that the results for $J=0$ and the $s$-wave magnet
are consistent with  Ref.~\onlinecite{Tserkovnyak2015}. 

\subsection{$J=0$}
For the TI surface without magnetism (i.e., $J=0$),
we have 
$\xi_{\pm} = \pm \lambda k - \mu$,
$h = \lambda k$,
$P_1 = -\lambda^2 k^2$,
$P_3 = 1$, and $P_0=P_2 = 0$,
which leads to
\begin{align}
G^{(0)}_{\alpha} = 0,
\end{align}
and
\begin{align}
G^{(1)}_{\alpha} = \frac{\alpha}{4 \lambda k}.
\end{align}
Therefore, we have
\begin{align}
\chi_{\rm OZ}
=& -\frac{4\lambda^2|e| \mu_{\rm B}}{\hbar}
\int\frac{d^2 k}{(2\pi)^2}
\left[\sum_{\alpha = \pm} \frac{\alpha}{4\lambda k}
f^\prime(\alpha \lambda k -\mu)
\right] \notag \\
= &\frac{|e|\mu_{\rm B}}{2\pi \hbar}
\left[2f(-\mu)-1\right].
\end{align}
For $T \rightarrow 0$, we have $f(-\mu) \rightarrow \frac{1+\mathrm{sgn}(\mu)}{2}$ (if $|\mu| \neq 0$), thus
$\chi_{\rm OZ} \rightarrow \mathrm{sgn}(\mu) 
\frac{|e|\mu_{\rm B}}{2\pi \hbar}$.

\subsection{$s$-wave magnet}
For the $s$-wave magnet (i.e., the ordinary ferromagnet),
where $X= 1$, we have
 $h = \sqrt{\lambda^2 k^2 + J^2}$
 and $\xi_{\pm } = \pm \sqrt{\lambda^2 k^2 + J^2} - \mu$.
 We then have 
 \begin{align}
\chi_{\rm OZ}
=& -\frac{4\lambda^2|e| \mu_{\rm B}}{\hbar}\notag \\
\times&\int\frac{d^2 k}{(2\pi)^2}
\left[\sum_{\alpha = \pm} \frac{\alpha}{4 \sqrt{\lambda^2 k^2 + J^2}}
f^\prime(\alpha\sqrt{\lambda^2 k^2 + J^2} -\mu)
\right] \notag \\
=& \frac{|e|\mu_{\rm B}}{2\pi \hbar} 
 \left[f(J-\mu)+f(-J-\mu)- 1\right].
\end{align}
For $T\rightarrow 0$, we 
have $\chi_{\rm OZ} \rightarrow 
\frac{|e|\mu_{\rm B}}{2\pi \hbar}[\Theta(\mu-J)+\Theta(J+\mu)- 1]
$.

\subsection{$p$-wave magnet \label{app:_ti_p}}
For the $p$-wave magnet,
where $X=k \cos \theta$, we have
 $h_{\bm{k}} = k \sqrt{\lambda^2 + J^2 \cos ^2 \theta}$
 and $\xi_{\pm } = \pm k \sqrt{\lambda^2 + J^2 \cos ^2 \theta} - \mu$.
 We then have 
 \begin{align}
\chi_{\rm OZ}
=& -\frac{4\lambda^2|e| \mu_{\rm B}}{\hbar}
\int\frac{d^2 k}{(2\pi)^2}
\left[\sum_{\alpha = \pm} \frac{\alpha}{4 h_{\bm{k}}}
f^\prime(\alpha h_{\bm{k}} -\mu)
\right] \notag \\
=& -\frac{\lambda^2|e| \mu_{\rm B}}{(2\pi)^2\hbar}
\int_0^{2\pi} d\theta \left\{ \frac{1}{\lambda^2 + J^2 \cos ^2\theta} \right. \notag \\
&\left. \int_0^{\infty} dk  \left[ \left(\frac{\partial f (h_{\bm{k}}-\mu) }{\partial k}\right) + \left(\frac{\partial f(-h_{\bm{k}}-\mu) }{\partial k}\right)  \right] \right\}
\notag \\
=& \frac{|e|\mu_{\rm B}}{2\pi \hbar}
\frac{\lambda}{\sqrt{\lambda^2 + J^2 }}
\left[2f(-\mu)- 1\right].
\end{align}
Note that we have used 
\begin{align}
\left(\frac{\partial f (\alpha h_{\bm{k}}-\mu) }{\partial k}\right) 
=& \alpha \frac{\partial h_{\bm{k}}}{\partial k} 
f^\prime(\alpha h_{\bm{k}}-\mu)\notag \\
=& \alpha \sqrt{\lambda^2 + J^2 \cos^2 \theta}
f^\prime(\alpha h_{\bm{k}}-\mu),
\end{align}
and 
\begin{align}
\int_0^{2\pi} \frac{d \theta}{\lambda^2 + J^2 \cos ^2\theta}
= \frac{2\pi}{\lambda \sqrt{\lambda^2 + J^2}}.
\end{align}
For $T\rightarrow 0$,
we have
$\chi_{\rm OZ} \rightarrow
\mathrm{sgn}(\mu)\frac{|e|\mu_{\rm B}}{2\pi \hbar}\frac{\lambda}{\sqrt{\lambda^2 + J^2}}$.

\subsection{$d$-wave magnet \label{app:ti_d}}
For the $d$-wave magnet,
where $X=k^2 \sin 2 \theta$, 
we have
 $h_{\bm{k}} = \sqrt{\lambda^2 k^2 + J^2 k^4 \sin^2 2 \theta}$
 and $\xi_{\pm } = \pm \sqrt{\lambda^2 k^2 + J^2 k^4 \sin^2 2 \theta} - \mu$.
 In this case, the analytical expression is obtained only when $T=0$,
 so we focus on this case.
 Before proceeding further, we introduce the Fermi wave number
 that satisfies $h_{\bm{k}} =|\mu|$,
 \begin{align}
 k_{\rm F} (\theta)= \sqrt{\frac{\sqrt{\lambda^4+4\mu^2 (J^2\sin ^2 2\theta)}-\lambda^2}{2(J^2\sin ^2 2\theta)}}.
 \end{align}
Note that $k_{\rm F} (\theta) \rightarrow \frac{|\mu|}{\lambda}$
for $\sin 2 \theta \rightarrow 0$.

Turning to $\chi_{\rm OZ}$ at $T=0$, 
we have
\begin{align}
\chi_{\rm OZ}
=& -\frac{4\lambda^2|e| \mu_{\rm B}}{\hbar}
\int\frac{d^2 k}{(2\pi)^2}
\left[\sum_{\alpha = \pm} \frac{\alpha}{4 h_{\bm{k}}}
f^\prime(\alpha h_{\bm{k}} -\mu)
\right] \notag \\
=& \mathrm{sgn}(\mu)\frac{\lambda^2|e| \mu_{\rm B}}{ (2\pi)^2 \hbar}
\int_0^{2\pi} d\theta \int_0^{\infty} dk
\frac{k}{h_{\bm{k}}} \delta(h_{\bm{k}}-|\mu|) \notag \\
=& \mathrm{sgn}(\mu)\frac{\lambda^2|e| \mu_{\rm B}}{ (2\pi)^2 \hbar}
\int_0^{2\pi} d\theta 
\left[
\frac{k}{h_{\bm{k}}} 
\left(\frac{\partial h_{\bm{k}}}{\partial k}\right)^{-1}
\right]
_{k=k_{\rm {F}}(\theta)} \notag \\
=& \mathrm{sgn}(\mu)\frac{\lambda^2|e| \mu_{\rm B}}{ (2\pi)^2 \hbar}
\int_0^{2\pi} d\theta 
\frac{1}{\sqrt{1+ 4\frac{J^2 \mu^2}{\lambda^4} \sin^2 2\theta}} \notag \\
=& \frac{|e|\mu_{\rm B}}{\hbar}\mathrm{sgn}(\mu)
\frac{K(w)}{\pi^2\sqrt{1+w}},
\end{align}
where $w$ and $K(x)$
are given as defined in the main text.
Note that the jump at $\mu = 0$
is the same value as 
that of $J=0$, since
$K(0) = \frac{\pi}{2}$.

\so{
\section{Detailed calculations of $\chi_{\rm OZ}$ using the orbital magnetization}
In this section, 
we present the detailed calculation of $\chi_{\rm OZ}$ using the orbital magnetization.
%we discuss the quantized jump in $\chi_{\rm OZ}$ in the viewpoint of the Berry curvature. 
First, we consider $s$-wave order parameter case. Next, we discuss $J=0$ case by taking $J\to 0$ limit.
\label{app:TI_interp}
}

\so{
\subsection{Basic properties of Hamiltonian }
The Hamiltonian in the momentum space under a Zeeman field reads
\begin{align}
 H_{\bm {k}} = 
 \begin{pmatrix}
     \Delta & \lambda (-k_y-ik_x) \\
     \lambda (-k_y+ik_x) & -\Delta
 \end{pmatrix},
\end{align}
with $\Delta=J+\mu_{\rm B} B \, (\Delta>0)$.
This Hamiltonian has particle-hole symmetry:
\begin{equation}
    CH_{\bm k}C^{-1}=-H_{-\bm k},
\end{equation}
with $C=\sigma_x K$, where $K$ is the complex conjugation operator.
The wave functions for the valence and conduction bands are given as
\begin{align}
    \ket{\psi^-_{\bm k}}=
    \begin{pmatrix}
        \soo{\sin} \frac{\eta_{\bm k}}{2} \\ -i e^{i\theta_{\bm k}}\soo{\cos} \frac{\eta_{\bm k}}{2}
    \end{pmatrix},
    \qquad \text{for} \quad E^-_{\bm k}=-E_{\bm k}
\end{align}
and
\begin{align}
    \ket{\psi^+_{\bm k}}= 
    \begin{pmatrix}
        -i e^{-i\theta_{\bm k}}\soo{\cos} \frac{\eta_{\bm k}}{2} \\ \soo{\sin} \frac{\eta_{\bm k}}{2}
    \end{pmatrix},
    \qquad \text{for} \quad E^+_{\bm k}=E_{\bm k},
\end{align}
where $E_{\bm k}=\sqrt{\lambda^2 k^2+\Delta^2}$, $\tan \eta_{\bm k}=\frac{\lambda k}\Delta\, (0\leq\eta_{\bm k}<\frac{\pi}{2})$, and $\theta_{\bm k}$ is the polar angle corresponding to $(k_x,k_y)\, (0\leq \theta_{\bm k}<2\pi)$.
We confirm $C\ket{\psi^+_{\bm k}}=\ket{\psi^-_{-\bm k}}$ and $C\ket{\psi^-_{\bm k}}=\ket{\psi^+_{-\bm k}}$.
The Berry curvatures for the valence and conduction bands are
\begin{align}
    \Omega^\pm_z({\bm k}) =i \left( \Braket{\frac{\partial \psi^\pm_{\bm k}}{\partial k_x} | \frac{\partial \psi^\pm_{\bm k}}{\partial k_y}} - \text{c.c.} \right) 
    =\mp\frac{1}{2} \frac{\lambda^2 \Delta}{E_{\bm k}^3}.
\end{align}
%and
%\begin{align}
%    \Omega^+_z({\bm k})& =i \left( \Braket{\frac{\partial \psi^+_{\bm k}}{\partial k_x} | \frac{\partial \psi^+_{\bm k}}{\partial k_y}} - \text{c.c.} \right) 
%    =-\frac{1}{2} \frac{\lambda^2 \Delta}{E_{\bm k}^3},
%\end{align}
%respectively.
}

\so{
We note that the $d$-wave case is also particle-hole symmetric but $p$-wave case is not. 
}

\so{
\subsection{Calculation of $\chi_{\rm OZ}$}
The integral of \eqref{eq:Mz} is performed as follows.
\begin{align}
    M_z=\frac{|e|\mu V}{2\pi\hbar} \sum_{\pm} \int_0^\infty dk\, k\,  \Omega^\pm_{\bm z}({\bm k})\Theta(\mu-E^\pm_{\bm k}).
\end{align}
}

\so{
(i) $\mu<-|\Delta|$ case:
}

\so{
\begin{align}
    M_z=\frac{|e|\mu V}{2\pi\hbar}  \int_{\sqrt{\mu^2-\Delta^2}/\lambda}^\infty \frac{k \lambda^2 \Delta }{2(\lambda^2 k^2+\Delta^2)^{3/2}} dk=-\frac{|e|\Delta V}{4\pi\hbar}.
\end{align}
Substituting $\Delta\to J+\mu_{\rm B} B$ and expanding with respect to $B$, we obtain
\begin{equation}
    \frac{1}{2}\chi_{\rm OZ}=\frac{1}{V}\frac{\partial M_z}{\partial B}=-\frac{|e|\mu_{\rm B} }{4\pi\hbar}.
\end{equation}
}

\so{
(ii) $-|\Delta|<\mu<|\Delta|$ case:
\begin{align}
    M_z=\frac{|e|\mu V}{2\pi\hbar}  \int_0^\infty \frac{k\cdot \lambda^2\Delta }{2(\lambda^2 k^2+\Delta^2)^{3/2}} dk=\frac{|e|\mu V}{4\pi\hbar}.
\end{align}
Therefore, 
\begin{equation}
    \frac{1}{2}\chi_{\rm OZ}=\frac{1}{V}\frac{\partial M_z}{\partial B}=0.
\end{equation}
}

\so{
(iii) $|\Delta|<\mu$ case:
}

\so{
\begin{align}
    M_z=& \frac{|e|\mu V}{2\pi\hbar} \biggl(\frac{1}{2}+ \int^{\sqrt{\mu^2-\Delta^2}/\lambda}_0 \frac{-k \cdot \lambda^2\Delta }{2(\lambda^2 k^2+\Delta^2)^{3/2}} dk  \biggr) \notag \\
    =&\frac{|e|\Delta V}{4\pi\hbar}.
\end{align}
Substituting $\Delta\to J+\mu_{\rm B} B$ and expanding with respect to $B$, we obtain
\begin{equation}
    \frac{1}{2}\chi_{\rm OZ}=\frac{1}{V}\frac{\partial M_z}{\partial B}=\frac{|e|\mu_{\rm B} }{4\pi\hbar}.
\end{equation}
}

\so{
In summary, $\chi_{\rm OZ}$ is given as
\begin{equation}
    \chi_{\rm OZ}=
        \frac{|e|\mu_B}{2\pi \hbar} {\rm sgn}(\mu) \Theta(|\mu|-J),
\end{equation}
which is exactly the same as the value obtained in the main text at $T=0$.
Finally, taking the $J\to 0$ limit, the $J=0$ case is reproduced.
These results show that the Berry curvature is responsible for the quantization of the jump observed in the $J=0$ and $s$-wave order parameter cases. 
}

\bibliography{OZ_Altermag}
\end{document}